\documentclass[amssymb,prl,twocolumn,floats]{revtex4}
\usepackage{bm}
\usepackage{graphicx}

\begin{document}
\preprint{November 12, 2001}

\title{Light and electric field control of ferromagnetism in
magnetic quantum structures}

\author{H. Boukari$^{(a)}$,P. Kossacki$^{(a,b)}$, M. Bertolini$^{(a)}$, D. Ferrand$^{(a)}$,
J. Cibert$^{(a)}$,\\ S. Tatarenko$^{(a)}$, A. Wasiela$^{(a)}$, J.A. Gaj$^{(b)}$, and T.
Dietl$^{(a,c)}$}

\affiliation{$^{(a)}$Laboratoire de Spectrom\'etrie Physique,
Universit\'e Joseph Fourier Grenoble 1 - CNRS (UMR 55~88),\\
Bo\^{\i}te Postale 87, F-38402 Saint Martin d'H\`eres Cedex,
France\\ $^{(b)}$Institute of Experimental Physics, Warsaw
University, Ho\.za 69, PL-00-681 Warszawa, Poland\\
$^{(c)}$Institute of Physics, Polish Academy of Sciences, PL-02-668 Warszawa, Poland}
\date{\today}

\begin{abstract}
A strong influence of illumination and electric bias on the Curie temperature and saturation value
of the magnetization is demonstrated for semiconductor structures containing a modulation-doped
p-type Cd$_{0.96}$Mn$_{0.04}$Te quantum well placed in various built-in electric fields. It is
shown that both light beam and bias voltage generate an isothermal and reversible cross-over
between the paramagnetic and ferromagnetic phases, in the way that is predetermined by the
structure design. The observed behavior is in quantitative agreement with the expectations for
systems, in which ferromagnetic interactions are mediated by the weakly disordered two-dimensional
hole liquid.
\end{abstract}


\draft



%\preprint

\maketitle

Soon after the discovery of carrier-controlled ferromagnetism in Mn-doped III-V \cite{Ohno92} and
II-VI \cite{Haur97} semiconductor compounds, it has become clear that these systems offer
unprecedented opportunities to exploit the powerful methods developed  for tuning carrier densities
in semiconductor quantum structures, in order to control the magnetic characteristics in these
systems \cite{Haur97,Kosh97,Ohno98,Koss00,Ohno00}. Such a control opens new prospects for
information storage and processing, as well as it makes it possible to examine the behavior of
strongly correlated systems as a function of externally controllable parameters. In the case of
III-V magnetic semiconductors, Koshihara {\it et al.} \cite{Kosh97} detected an enhancement of
ferromagnetism by illumination of an (In,Mn)As/GaSb heterostructure, an effect assigned to the
presence of an interfacial electric field that drives the photo-holes to the magnetically active
(In,Mn)As layer. More recently, Ohno {\sl et al.} \cite{Ohno00} demonstrated that a gate voltage of
$\pm 125$ V changes the Curie temperature $T_{\mbox{\small{C}}}$ by about 1 K in a field-effect
transistor structure containing an (In,Mn)As quantum well (QW).

In the case of II-VI diluted magnetic semiconductors (DMS), Mn does not introduce any carriers.
Hence, hole-mediated ferromagnetic interactions can be induced by modulation-doping of
heterostructures \cite{Diet97}. Due to the valence band structure, $T_{\mbox{\small{C}}}$ is
typically lower in II-VI than in III-V DMS. At the same time, however, it may be expected
\cite{Haur97,Koss00} that, owing to the small background hole density, the strength of the carrier
mediated ferromagnetic interactions can be tuned over a wider range in II-VI than in III-V DMS.

In this paper, we present photoluminescence (PL) studies of modulation-doped p-type (Cd,Mn)Te QW.
The (Cd,Zn,Mg)Te barriers are doped either p- or n-type, so that p-i-p or p-i-n structures are
formed. The QW in these systems are ferromagnetic below about 3~K. We show that, depending of the
sample layout, the ferromagnetism is either destroyed or enhanced during illumination by photons
with energy greater than the band gap of the barrier material. In both cases, the switching
process is isothermal and reversible. Moreover, we demonstrate that the reverse biasing of the
p-i-n diode by a voltage smaller than 1~V turns the ferromagnet into a paramagnetic material.
Importantly, this strong effect of light and electric field can be readily explained by
considering the distribution of carriers and photo-carriers in the p-i-p and p-i-n structures. At
the same time, by tracing how the system properties vary with the hole density in the QW, we
identify processes accounting for the magnitude of the Curie temperature and spontaneous
magnetization in this low-dimensional ferromagnetic system.

Our samples were grown coherently onto a Cd$_{0.88}$Zn$_{0.12}$Te substrate by molecular beam
epitaxy, exploiting previous expertise in p-type \cite{Arno99} and n-type \cite{Arno00} modulation
doping of tellurides. As shown in Fig.~1, the structures contain a single 8 nm QW of (Cd,Mn)Te,
with typically 3\% to 5\% Mn and (Cd,Mg,Zn)Te barriers.  In the back barrier of the p-i-p
structure, a nitrogen-doped p-type layer is inserted at 100~nm from the QW. A hole gas in the QW
is created by doping the top barrier with nitrogen acceptors of concentration exceeding
$10^{17}$~cm$^{-3}$. The corresponding spacer layer thickness is 20~nm. In the p-i-n diode, the
back barrier doped with aluminum (n-type) resides 320~nm away from the QW, and the spacer between
the QW and the p-doped layer is diminished to 10~nm. This leads to a hole density in the QW of
about $2\times 10^{11}$ cm$^{-2}$ in both p-i-p and p-i-n structures, so that the carriers occupy
only the ground-state heavy-hole subband. A semi-transparent gold film is evaporated on top of the
p-i-n diodes, and then $2\times2$ mm$^2$ squares are formed by Ar-ion etching down to the n-type
layer, a procedure followed by the deposition of In contacts. In these diodes, non-linear
current-voltage characteristics are observed up to room temperature.

\begin{figure}
\includegraphics*[width=85mm]{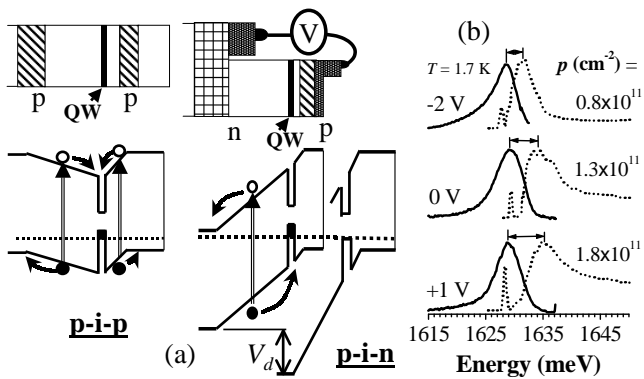}
\caption[]{(a)Layout of p-i-p and p-i-n structures containing a (Cd,Mn)Te quantum well, and band
alignments with and without bias voltage $V_d$, showing the expected migration of the
photocarriers. (b) PL (solid line) and PLE (dashed line) showing the Moss-Burstein shift (arrows),
$versus$ bias voltage $V_d$, on a p-i-n structure containing a Cd$_{0.995}$Mn$_{0.005}$Te QW. The
low energy feature in PLE is the laser peak.}
 \label{fig:CdMnTe_1}
\end{figure}

Following the established procedure \cite{Haur97,Koss00,Gaj94,Koss99}, we probe the properties of
the system by PL and its excitation (PLE).  Owing to the exchange coupling between the band
carriers and the Mn spins, the PL line splitting is proportional to the Mn magnetization.This
allows us to measure the magnetization locally, in a single QW, with a much higher sensitivity than
using a conventional magnetometer. PL is excited by an Al$_2$O$_3$:Ti or HeNe laser, whose photon
energy is below the barrier gap, and the output power is kept below 9~mW/cm$^2$. An additional
above-barrier illumination, which aims at affecting the ferromagnetism (by increasing or reducing
the hole density, as explained below), is provided either by a halogen lamp screened by one blue
and a variable number of gray filters, or by an Ar-ion laser. The Moss-Burstein shift between the
PL and PLE lines serves us to determine the hole density \cite{Koss99} at given values of
illumination intensity and bias voltage. To increase the accuracy, these measurements were also
carried out in a magnetic field over 1~T, in which the hole gas is entirely spin polarized, so that
the magnitude of the shift is doubled. Fig.1b shows the influence of the bias voltage $V_d$, on the
Moss-Burstein shift (arrows) for a sample with only 0.5\% Mn in the QW so that no ferromagnetic
ordering is observed. This procedure appears to provide an accurate evaluation of the relative hole
density, whereas its absolute value is determined within a factor of 2. It was checked that for the
depleted QW both spectral position and giant Zeeman splitting of the exciton reflectivity line are
consistent with those expected for the nominal values of QW width and Mn concentration (3\% to
5\%). Furthermore, an increase of the spin temperature \cite{Koni00} can be deduced from the Zeeman
splitting but only for much higher illumination intensities than in the present study.

Figure 2 presents PL spectra collected for p-i-p and p-i-n structures in the absence of an
external magnetic field at various temperatures, illumination intensities, and bias voltages. The
PL line corresponds to the e1$\rightarrow$ hh1 transitions. Its splitting, accompanied by a red
shift of its lower component \cite{Haur97,Koss00}, signals the transition to an ordered phase. The
splitting energy is proportional to the spontaneous magnetization within a magnetic domain. Since
the easy axis is oriented along the growth direction, the emitted light contains two circularly
polarized components of equal intensity corresponding to the two possible orientations of the
magnetic domains. Applying a small field in the Faraday geometry results in variations of the
splitting and circular polarization of the lines in agreement with the evolution expected from
such domains \cite{Koss00}. As shown, the phase transition occurs not only on lowering the
temperature [Figs.~2(a) and 2(c)], but also isothermally, on changing the illumination [Fig.~2(b)]
or bias voltage [from Fig.~2(c) to Fig.2~(d)]. Remarkably, while above barrier illumination
destroys ferromagnetism in the p-i-p structures [Fig.~2(b)], it enhances the spontaneous
magnetization in the p-i-n diodes [dotted line in Fig.~2(c)]. Since in the systems in question
ferromagnetic interactions are mediated by holes, we assign the observed behavior to the influence
of illumination and bias voltage on the hole density in the QW which contains the Mn spins. To put
these considerations onto a more quantitative basis we will first relate the illumination
intensity and bias voltage to the hole density, and then examine how the latter determines
$T_{\mbox{\small{C}}}$ and spontaneous magnetization.

\begin{figure}
%\widetext
\includegraphics*[width=80mm]{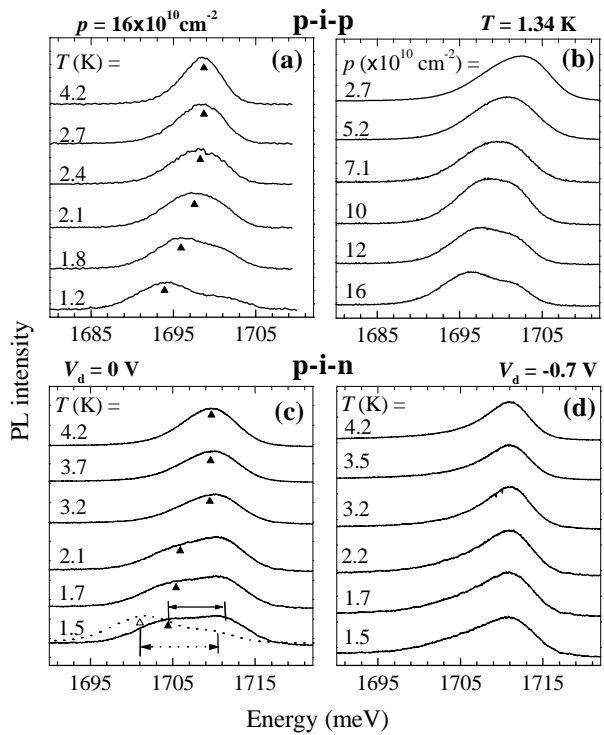}
\caption[width=60mm]{PL spectra for a modulation-doped p-type Cd$_{0.96}$Mn$_{0.04}$Te QW located
in a p-i-p structure and a modulation-doped p-type Cd$_{0.95}$Mn$_{0.05}$Te QW in a p-i-n diode;
(a) p-i-p structure without additional illumination (i.e., constant hole density) at various
temperatures; (b) same p-i-p structure with above barrier illumination (to reduce the hole
density) at fixed temperature; (c) p-i-n structure without bias at various temperatures: the hole
density is constant (solid lines) or increased (dotted line) by additional Ar-ion laser
illumination; (d) p-i-n structure with a -0.7 V bias (depleted QW) at various temperatures.
Splitting and shift of the lines mark the transition to the ferromagnetic phase.}
 \label{fig:CdMnTe_2}
\end{figure}
%\end{widetext}
%\twocolumn

As shown for similar p-i-p structures with a lower Mn content \cite{Koss99}, photons of energy
larger than the barrier bandgap reduce the hole density in the QW. According to Fig.~1, the
photo-electrons created in this way migrate to the QW, where they recombine with the pre-existing
holes. At the same time, photo-holes are trapped in the barrier layers, and tunnel rather slowly
back to the QW. The interplay of these processes determines the steady hole density. In agreement
with the previous studies \cite{Koss99}, the relation between the light intensity $I$ and the hole
density $p$ determined from the Moss-Burstein shift is well described by the formula expected for
tunneling through the triangular barrier arising due to the electric field induced by the presence
of the 2D hole gas in the QW, $I = A(p-p_0)\exp(-\beta \sqrt{p})$, where $p_0$ is the hole density
without illumination, and the parameters $A$ and $\beta$ are proportional to the rates of electron
migration and hole tunneling, respectively. We find their values to be similar to those determined
previously \cite{Koss99}, {\em e.g.}, $\beta = 3.34\times 10^7$~m. Moreover, photo-carriers created
directly by PL excitation into the QW , recombine fast so that they do not affect the carrier
density \cite{Koss99}.

Turning to the p-i-n diodes, the band structure (Fig.~1), and the fact that the penetration length
of light is larger than one micrometer, show that the built-in electric field drives the majority
of the photo-holes towards the QW. This causes an enhancement of ferromagnetism under
illumination, an effect visible in Fig.~2(c). In addition, biasing the p-i-n diodes controls the
QW hole density: the density, evaluated from the Moss-Burstein shift, changes from $p_0 \approx
2\times 10^{11}$~cm$^{-2}$ at zero bias down to zero at $-0.7$~V. Accordingly, at this bias, no
signature of ferromagnetism is observed in the PL spectra down to 1.5~K, as shown in Fig.~2(d).
Knowing the sample layout, we can estimate the bias $V_d$, which has to be applied in order to
deplete the QW entirely. The main contribution is that of the electric field across the
capacitance formed by the QW and the n-type layer. For the dielectric constant of CdTe, $\epsilon
=10$, this leads to $V_d \approx ep_0L/\epsilon_o\epsilon =1$~V which reproduces well the
experimental value.

An important aspect of our data is that they provide detailed information on the dependence of
$T_{\mbox{\small{C}}}$ and the spontaneous magnetization on the hole density $p$. There are two
ways of crossing the phase boundary in our system: by varying the temperature $T$ at constant $p$,
or by varying $p$ at constant $T$. The experimental values of $T_{\mbox{\small{C}}}$ have been
obtained by tracing the position of the lower PL line as a function of $T$ at constant $p$, as
shown in Fig.~3(a). This procedure avoids uncertainties associated with the renormalization of the
PL energy by carrier-carrier correlation, which depends more on $p$ than on $T$, an effect visible
in Fig.~2(b). A mean-field model for $T_{\mbox{\small{C}}}$ in low-dimensional structures has been
derived by some of the present authors \cite{Haur97,Diet97} and others \cite{Lee00,Jung01}. The
applicability of the mean-field approximation (MFA) is justified by the long-range character of
the ferromagnetic interactions in question. In the theory adopted here \cite{Haur97,Diet97} the
contribution of short range antiferromagnetic interactions is included in terms of an effective
spin density and with an effective Curie-Weiss temperature which is negative (noted
$-T_{\mbox{\small{AF}}}$ hereafter) as measured for undoped DMS.  As a result, the Curie-Weiss
temperature is given by $T_{\mbox{\small{C}}} =T_{\mbox{\small{F}}}-T_{\mbox{\small{AF}}}$, where
$T_{\mbox{\small{F}}}$ is proportional to the Pauli susceptibility of the hole liquid
$\tilde{\chi}_h$. Hence in a degenerate gas, $\tilde{\chi}_h$ is proportional to the density of
states at Fermi energy, which is independent of $p$ for an ideal 2D gas \cite{Diet97}. The
hole-hole interactions are incorporated into $\tilde{\chi}_h$ according to the Fermi liquid
theory.
 In the following, when evaluating $\tilde{\chi}_h$ at such low values of the hole
density \cite{Koss00}, the assumption that the hole liquid is in the highly degenerate limit is
relaxed, and the electrostatic disorder is taken into account by a gaussian broadening of the hole
density of states. Then, $T_{\mbox{\small{C}}}$ evolves smoothly from $-T_{\mbox{\small{AF}}}$ to
the constant value pertaining to the clean degenerate 2D liquid.

 As shown in Fig.~3(b), there is a good agreement between experimental and
theoretical values of $T_{\mbox{\small{C}}}$ as a function of the hole density. The theoretical
curve is drawn with two adjustable parameters: the Fermi liquid parameter $A_F = 2.1$ and the
full-width at half maximum of the density-of-states gaussian broadening $\Gamma = 1.7$~meV. Such a
two-fold enhancement of the Pauli susceptibility by the interactions, as implied by the magnitude
of $A_F$, is to be expected for the hole density range in question \cite{Jung01}. According to
Fig.~3(b), it is principally the non-zero value of $\Gamma$ which accounts for the lowering of
$T_{\mbox{\small{C}}}$ observed in the regime of small hole densities. Indeed, the value of
$\Gamma$ determined here compares favorably with the activation energy of 1.5~meV, determined for
positively charged excitons in a similar CdTe QW \cite{Brin99}: these charged excitons were
thought to be trapped by potential fluctuations brought about by acceptors residing at 50~nm from
the QW on its both sides.

  \begin{figure*}
\includegraphics*[width=140mm]{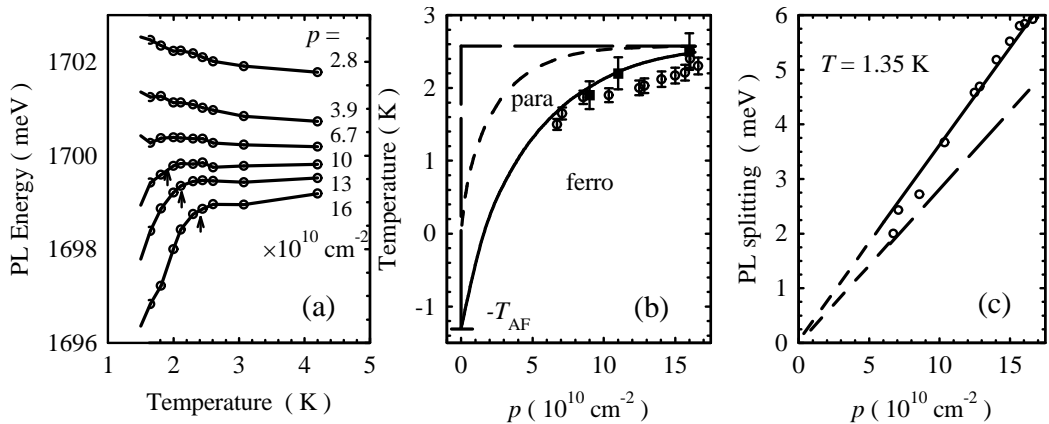}
\caption[width=160mm]{(a) Temperature dependence of the low-energy peak of the PL spectra for
selected values of the hole density (changed by illumination) for a p-type
Cd$_{0.96}$Mn$_{0.04}$Te QW. Arrows indicate the points taken as indicating the critical
temperature of the ferromagnetic transition ($T_{\mbox{\small{C}}}$). (b) $T_{\mbox{\small{C}}}$
$versus$ hole density for two samples (circles, same as (a), and squares). The line shows
$T_{\mbox{\small{C}}}$ values calculated within the mean field model and for various assumptions
about the hole spin susceptibility: the dashed line is for a 2D degenerate Fermi liquid, the
dotted line takes into account the effect of non-zero temperature on a clean Fermi liquid, and the
solid line is obtained assuming a gaussian broadening of the density-of-states. (c) Zero-field
splitting of the PL line at 1.35~K, as a function of the hole density. The solid straight line is
drawn through the data points; the dashed line is calculated within the mean-field model
neglecting the effects of hole-hole correlation on the optical spectra. Both lines cross zero as
shown by the dotted lines} \label{fig:CdMnTe_3}
  \end{figure*}

%\twocolumn
It has been shown \cite{Diet97} that, in the ideal 2D case, the hole gas should be completely
polarized immediately below $T_{\mbox{\small{C}}}$. Therefore, the magnetization of the Mn spins
induced  by the molecular field of the spin polarized holes is expected to increase linearly with
the hole density. Experimental values of the zero-field splitting at 1.35~K, plotted versus the
hole density in Fig.~3(c), corroborate this expectation in the density range where the zero-field
splitting is observed. The small difference between the experimental values and the results of the
mean-field model \cite{Diet97} may suggest a possible influence of carrier-carrier correlation on
optical spectra. However, the limited accuracy of our determination of the absolute value of the
hole densities precludes any definitive conclusion. Nevertheless, it is worth noting that this
linear dependence is markedly different from that found for (Zn,Mn)Te:N epilayers \cite{Ferr01},
in which a part of the Mn spins escape the ferromagnetic interaction at low carrier density,
probably due to hole localization in the presence of acceptors in the DMS.

In conclusion, our results show how manipulating the density $p$ of the two-dimensional hole liquid
affects the ferromagnetic properties of magnetic quantum wells. In particular, in agreement with
the theoretical expectations \cite{Diet97}, the spontaneous magnetization is linear in $p$. While
no dependence of $T_{\mbox{\small{C}}}$ on $p$ is expected for a fully degenerate ideal 2D hole
liquid, we observe a dependence which is rather weak and well explained by disorder effects.
Moreover, our findings show that both photon beam and electric field can isothermally drive the
system between the ferromagnetic and paramagnetic phases, in a direction which can be selected by
an appropriate design of the structure. This offers new tools for patterning magnetic
nanostructures as well as for information writing and processing, beyond the heating effects of
light exploited in the existing magneto-optical memories. Obviously, however, practical application
of the tuning capabilities put forward here have to be preceded by a progress in the synthesis of
functional room temperature ferromagnetic semiconductors. As far as II-VI compounds are concerned,
according to theoretical suggestions \cite{Diet00,Sato01}, structures containing ZnO, such as
p-type (Zn,Mn)O/(Zn,Mg)O, appear to be a prospective material system.

The French-Polish collaboration is supported by the Polonium program. The authors thank Y.~Genuist
and R.~Hammelin for technical assistance.

\end{document}